\documentclass[12pt]{article}
\usepackage{color}
\newcommand{\beq}{\begin{equation}}
\newcommand{\eeq}{\end{equation}}
\newcommand{\beqa}{\begin{eqnarray}}
\newcommand{\eeqa}{\end{eqnarray}}
\newcommand{\beqar}{\begin{eqnarray*}}
\newcommand{\eeqar}{\end{eqnarray*}}

\newcommand{\al}{\alpha}
\newcommand{\be}{\beta}

\def\non          {\nonumber}
\def\ha           {\mbox{$\frac{1}{2}$}}

\def\Tr           {\mbox{\rm Tr}\,}

\def\cd           {{\cdot}}
\def\ran          {\rangle}
\def\lan          {\langle}
\def\fsk    {k\!\!\!\!/\,}
\def\fsH    {H\!\!\!\!/\,}

\newcommand{\ga}{\gamma}

\newcommand{\lam}{\lambda}

\newcommand\bPsi{{\bar \Psi }}

\newcommand{\z}{\zeta}

\newcommand{\labell}[1]{\label{#1}} 
\newcommand{\reef}[1]{(\ref{#1})}

\newcommand\veps{\varepsilon}

\newcommand\cD{{\cal D}}

\newcommand\bu{\bar{u}}

\newcommand{\dga}{\dot{\gamma}}
\newcommand{\dde}{\dot{\delta}}

\newcommand{\gab}{\bar{\gamma}}

\def\sst#1{{\scriptscriptstyle #1}}
\def\0{{\sst{(0)}}}
\def\1{{\sst{(1)}}}
\def\2{{\sst{(2)}}}
\def\3{{\sst{(3)}}}
\def\4{{\sst{(4)}}}
\def\5{{\sst{(5)}}}
\def\6{{\sst{(6)}}}
\def\7{{\sst{(7)}}}
\def\8{{\sst{(8)}}}

\begin{document}
\baselineskip 18pt%
\begin{titlepage}
\vspace*{1mm}%
\hfill
\vbox{

    \halign{#\hfil         \cr
           } 
      }  
\vspace*{9mm}
\vspace*{9mm}%

\center{ {\bf \Large  SYM, Chern-Simons, Wess-Zumino Couplings and their higher derivative
corrections  in IIA Superstring theory
}}\vspace*{1mm} \centerline{{\Large {\bf  }}}

\begin{center}
{Ehsan Hatefi \small $^{1,2,3,4}$}

\vspace*{0.6cm}{ {\it
{\small $^{1}$International Centre for Theoretical Physics, Strada Costiera 11, Trieste, Italy},
\vskip.1in
{\small $^{2}$ Centre for Research in String Theory, School of Physics and Astronomy,
Queen Mary University of London,Mile End Road, London E1 4NS, UK},
\vskip.06in

{\small $^{3}$National Institute for Theoretical Physics ,
School of Physics and Centre for Theoretical Physics,University of the Witwatersrand, Wits, 2050, South Africa}
\vskip.06in
 {\small $^{4}$E-mails:ehsan.hatefi@cern.ch,ehsanhatefi@gmail.com}

}}
\vspace*{0.1cm}
\vspace*{.1cm}
\end{center}
\begin{center}{\bf Abstract}\end{center}
\begin{quote}

We find the entire form of the amplitude of two fermion strings (with different chirality), a massless scalar field and one closed string Ramond-Ramond (RR) in IIA superstring theory which is different from its IIB one. We make use of a very particular gauge fixing and  explore several new couplings in IIA.  All infinite $u$- channel scalar poles and $t,s$- channel fermion poles are also constructed.  We find new form of higher derivative corrections to two fermion two scalar couplings and show that the first simple $(s+t+u)-$ channel scalar pole  for $p+2=n$ case can be obtained by having new higher derivative corrections to SYM couplings at third order of $\alpha'$. We find that the general structure and the coefficients of higher derivative corrections to two fermion two scalar couplings are completely different from the derived $\alpha'$ higher derivative corrections of type IIB.
\end{quote}
\end{titlepage}

\section{Introduction}

D$_p$-branes  (with $p$ as the spatial dimension of a  D$_p$-brane) are the fundamental/key objects in superstring theory\cite{Polchinski:1995mt,Witten:1995im,Polchinski:1996na}. It is known that a BPS brane does carry the so called Ramond-Ramond\footnote{From now on, we show it with RR.} charge.  We recommend several important papers to work with superstring perturbation theory \cite{Witten:2013cia}  or to deal with holomorphic string amplitudes \cite{Witten:2013tpa}. We also refer to several fascinating papers on mathematical structures and hidden symmetries of the scattering amplitudes \cite{D'Hoker:2013eea,Hatefi:2013hca,Arkani-Hamed:2013jha,Arkani-Hamed:2013kca}.

For various reasons, one needs to deal with branes' dynamics. In  an interesting paper \cite{Ademollo:1974fc}, diverse transitions of open/closed strings have been comprehensively explained. In order to highlight several dual prescriptions of branes, we introduce to the interested reader a review of string dualities \cite{Polchinski:1996nb}. For the completeness, let us just remind that some specific examples , such as $D0/D4$ system \cite{Hatefi:2012sy}, the description of world volume of BPS branes from super gravity side and eventually the realization of Ads brane world \cite{Hatefi:2012bp} are already shown.

\vskip.1in

If we are able to explore either bosonic or supersymmetric effective actions of BPS branes  at low energy limit by scattering approach, then we might hope very much  to indeed  learn about branes' dynamics. In Myers paper  \cite{Myers:1999ps} by taking into account several  D$_p$-brane configurations really  it is well explained how to employ bosonic actions. Although it is very difficult to completely discover supersymmetric effective actions \cite{Howe:2006rv}, in this paper we try to have progress in our understandings of these particular actions.
\vskip 0.1in

Having taken several important papers (to include an effective action of a single bosonic brane \cite{Leigh:1989jq} and its supersymmetrized version \cite{Cederwall:1996pv}), one may be able to follow what has been achieved.

 \vskip .2in

The aim of this paper is to show that the S-matrix of the mixed closed/open string amplitudes of type IIA is entirely different from type IIB and in fact several new couplings will come out from direct computations of conformal field theory techniques of type IIA . In addition to that we want to show that the derived corrections of type IIB do not work for type IIA.

  \vskip .01in

 Basically we have shown that in the computations of the S-matrix of one RR $(C)$, two fermions and one scalar of type IIB there are infinite u-channel scalar /gauge poles. While for the same amplitude in type IIA (with different chirality) we will see that all infinite u-channel gauge poles will be disappeared .
  \vskip .1in
 Another interesting point is that unlike IIB, here in IIA we do not have any $\alpha'$ corrections to two fermions, one gauge and one scalar field.

 \vskip .2in

 The very interesting point which comes out of long computations of $<V_{C}V_{\phi} V_{\bar\psi}V_{\psi}>$ of type IIA is that, not only the general structure of higher derivative corrections of two fermions- two scalars of IIA are different from IIB but also their coefficients are different.

We find these corrections at third order of $\alpha'$. It is also of high importance to focus on the point that several new couplings could be explored in type IIA of this paper.
 \vskip .2in

  To deal with effective field theory, one must know Myers terms, Wess-Zumino \cite{Hatefi:2010ik,Hatefi:2012zh} and the Chern-Simons actions.
    The Pull-back, the so called Taylor expansion and essentially all order Myers terms of BPS/non-BPS branes have been already constructed in \cite{Hatefi:2012wj}.
\vskip 0.02in

Some nice works in favor of the entire low energy actions have been carried out \cite{Koerber:2002zb,Keurentjes:2004tu,Denef:2000rj}. One can introduce some of the basic papers of mixed closed/open BPS branes where various motivations/applications to scattering theory/mathematical physics and similar results on  BPS branes have been involved in \cite{Hashimoto:1996bf}. It is also explained in  \cite{Hatefi:2012sy} that how to dissolve the branes with lower dimensions inside higher dimensional ones by taking some of the higher derivative $\alpha'$ corrections \cite{Hatefi:2012wz}. To be precise , if we take  $D(-1)/D3$ system then the description of its $N^2$
entropy could be explained , if and only if  $\alpha'$ higher derivative corrections of known Myers terms were derived. In \cite{Hatefi:2012wz,McOrist:2012yc} some arguments in favor of the  applications to  recent couplings are given.
\vskip 0.1in


    Keeping in mind either Myers terms or new WZ actions \cite{Hatefi:2010ik},\cite{Hatefi:2012ve,Hatefi:2012rx,Hatefi:2012cp,Garousi:2007fk,Hatefi:2008ab}, we are willing to find out new couplings and in particular new  higher derivative corrections of two fermions (with different chirality)-two scalars  up to third order of $\alpha'$  by producing the first simple  $(t+s+u)$-channel scalr pole (just for $p+2=n$ with $n$ becomes the index of RR field strength ) of our S-matrix $<V_{C} V_{\phi} V_{\bar\psi}V_{\psi} >$. These corrections might have some crucial rule in F-theory  \cite{Vafa:1996xn} or  M-theory \cite{Hatefi:2012sy,Hatefi:2012wz,McOrist:2012yc}.


\vskip.02in

Let us make some remarks. As we will observe in the explicit form of RR vertex operator (in ten dimensions of non compact space),  we did not include winding modes. Thus it is important to highlight that applying direct computations of CFT is much better than making use of T-duality to the old amplitudes of type IIB. We have argued in \cite{Hatefi:2013eia} that in order to be sure that all the terms of the related S-matrix are appeared in its final form and to derive correct form of higher derivative $\alpha'$ corrections  with exact coefficients , we should carry out direct CFT computations at each level of the amplitude.

\vskip.01in

Notice that For mixed open-closed amplitudes, computations should be made by taking  path integral method and all scalar/gauge /fermion (open) propagators should be derived by CFT techniques. Meanwhile it is discussed that in the case of  RR , one has to use both
$(\alpha_n,\tilde{\alpha}_n)$ oscillators. The authors in \cite{Billo:2006jm} explained that one can work out  both oscillators by dealing with open strings and RR could be taken by definition as open strings' composite states.

 To our knowledge this means that we can define background fields as some functions of SYM. It is also described in Myers paper \cite{Myers:1999ps} that  to make sense of all supergravity fields , we need to employ Taylor expansion to the effective field theory. It is also described in \cite{Park:2007mc} that the presence of BPS quantum effects of the open strings might shed new light on understanding  the host brane' curvature.

\vskip 0.06in

Here is the organization of this paper. First we make some points on the notations. Then we move on to explore the complete form of the $<V_{C} V_{\phi} V_{\bar\psi}^{\dga}V_{\psi}^{\dde} >$ of type IIA.  Then we expand the amplitude and make use of the low energy limit to be able to re construct all order vertices and we find out several new couplings in field theory of type IIA . These couplings can be discovered just by comparing them with this S-Matrix of type IIA. Finally we produce all $u$- channel scalar poles and $t,s$- channel fermion poles and make various comments   (for further details see Appendix A and B of \cite{Hatefi:2012wj,Hatefi:2012ve})

\section{ Notations and Analysis of  $<V_{\bar\psi}V_{\psi}V_{\phi}>$ }

The goal of this section is to provide necessary details to obtain the exact and complete form of a mixed and technically five point (physically four point) amplitude involving two fermions, a massless scalar and a closed string RR in the bulk  in type IIA superstring theory. To do so, of course we need to apply conformal field theory (CFT) techniques. The computations of this particular case in IIB has been done in a recent paper \cite{Hatefi:2013eia} but as we will see various things change in higher point functions of string theory. For the completeness we want to emphasize some of the string computations in various string theories \cite{Hatefi:2010ik,Hatefi:2012ve,Hatefi:2012rx,Hatefi:2014lva,Kennedy:1999nn,Chandia:2003sh}.
It is important to mention that the three point function including two fermions (with the same chirality ) and one gauge (scalar) field in IIB is done in \cite{Polchinski:1998aa} (\cite{Hatefi:2013eia}) and since the correlation function for two spin operators and a fermion field (the Wick-like rule) is not changed \cite{Liu:2001qa,Kostelecky:1986xg} , we expect that the same result for three point function is being held for IIA (with different chirality ) as well.

\vskip.1in

It is worth trying to address an important technical issue, namely to be able to have various S-matrices of a closed and several open strings (fermions and/or currents), the so called Wick-like rule should have been generalized. This generalization was carried out in \cite{Hatefi:2010ik,Hatefi:2012wj,Garousi:2007fk,Hatefi:2008ab}.

We just point out to the related vertex operators in IIA

\beqa
V_{\phi}^{(-1)}(x) &=&e^{-\phi(x)} \xi_{i}\psi^i(x)e^{\alpha' ik\cd X(x)}\nonumber\\
V_{\phi}^{(-2)}(y) &=&e^{-2\phi(y)}\xi_{i}\bigg(\partial
X^i(y)+\alpha' ik\cd\psi\psi^i(y)\bigg)e^{\alpha' ik\cd X(y)},
\nonumber\\
V_{\bPsi}^{(-1/2)}(x)&=&\bu^{\dga}e^{-\phi(x)/2}S_{\dga}(x)\,e^{\alpha'iq.X(x)}, \nonumber\\
V_{\Psi}^{(-1/2)}(x)&=&u^{\dde}e^{-\phi(x)/2}S_{\dde}(x)\,e^{\alpha'iq.X(x)}.
\label{d4Vs}
\eeqa
  where the on-shell conditions for scalar are $k^2=k.\xi=0$ and for the fermions are $q^2=q_a \gamma^a \bu^{\dga}= u^{\dde}q_b \gamma^b=0 $.
  $u^{\dde}$ should be regarded as  the wave function of  fermion and one could use charge conjugation matrix $C^{\alpha\be}$  to work with the spin indices. We suggest \cite{Hatefi:2013eia} for seeing the meanings of the traces and observing some other details.

\subsection{ The complete form of $\bigg( RR \bar\psi^{\dga} \psi^{\dde}\phi\bigg)$ in type IIA  }

In this section we are going to work with the world volume of BPS branes to be able to find out the comprehensive form to all orders of  $\alpha'$ of  the S-matrix of two fermions (with different chirality of RR ), a massless scalar field and a closed string RR. Having set that, this amplitude just made sense in IIA, thus all the corrections that we have found in IIB \cite{Hatefi:2013eia}can not be applied to IIA.

Let us clarify more details. We need to begin with the following vertices for this particular amplitude
\beqa
V_{\phi}^{(0)}(x) &=& \xi_{i}\bigg(\partial
X^i(x)+\alpha' ik\cd\psi\psi^i(x)\bigg)e^{\alpha' ik\cd X(x)}, \nonumber\\
V_{C}^{(-\frac{1}{2},-\frac{1}{2})}(z,\bar{z})&=&(P_{-}\fsH_{(n)}M_p)^{\al\be}e^{-\phi(z)/2}
S_{\al}(z)e^{i\frac{\alpha'}{2}p\cd X(z)}e^{-\phi(\bar{z})/2} S_{\be}(\bar{z})
e^{i\frac{\alpha'}{2}p\cd D \cd X(\bar{z})},
\label{1d4Vs}
\eeqa

The definitions of the RR's field strength in IIA and projection operator are
 \beqa
\fsH_{(n)} &=& \frac{a
_n}{n!}H_{\mu_{1}\ldots\mu_{n}}\ga^{\mu_{1}}\ldots
\ga^{\mu_{n}},n=2,4 ,a_n=i \nonumber\\
P_{-} &=& \ha (1-\ga^{11})
\nonumber\eeqa

It is also recommended to work with just some holomorphic functions as follows:
\begin{eqnarray}
\lan X^{\mu}(z)X^{\nu}(w)\ran & = & -\frac{\alpha'}{2}\eta^{\mu\nu}\log(z-w) , \non \\
\lan \psi^{\mu}(z)\psi^{\nu}(w) \ran & = & -\frac{\alpha'}{2}\eta^{\mu\nu}(z-w)^{-1} \ ,\non \\
\lan\phi(z)\phi(w)\ran & = & -\log(z-w) \ .
\labell{prop}\end{eqnarray}
that is why we apply
doubling tricks to our calculations \cite{Hatefi:2012wj}.

It is understood that the amplitude does not depend on the picture of BPS branes, and we prefer to carry out the computations in the following picture

\begin{eqnarray}
{\cal A}^{C \phi  \bar\psi\psi} & \sim & \int dx_{1}dx_{2}dx_{3}dzd\bar{z}\,
  \lan V_{\phi}^{(0)}{(x_{1})}
V_{\bar\psi}^{(-1/2)}{(x_{2})}V_{\psi}^{(-1/2)}{(x_{3})}
V_{RR}^{(-\frac{1}{2},-\frac{1}{2})}(z,\bar{z})\ran,\labell{sstring}\eeqa

  We just look for the ordering of  $\Tr(\lambda_1\lambda_2\lambda_3)$. If we think of the given vertices then we realize that the S-matrix has to be divided to two different parts. We first find its first part as below:

\vskip.01in

To this aim we have to discover  the correlator of four spin operators (in ten dimensions) with different chirality \cite{Friedan:1985ge,Hartl:2010ks} as follows
\beqa
  <S_{\alpha}(z_4)S_{\beta}(z_5)S^{\dga}(z_2)S^{\dde}(z_3)>= \bigg(\frac{x_{45}x_{23}}{x_{42}x_{43}x_{52}x_{53}}\bigg)^{1/4}
  \bigg[\frac{C_{\alpha}^{\dde}C_{\beta}^{\dga}} {x_{43}x_{52}}-\frac{C_{\alpha}^{\dga}C_{\beta}^{\dde}} {x_{42}x_{53}}+\frac{1}{2}\frac{(\gamma^\mu C)_{\alpha\beta}\,(\gab_\mu C)^{\dga\dde}} {x_{45}x_{23}}\bigg]\label{ham}\eeqa

the next step is to actually substitute the above correlator  inside the amplitude  and one can read off its S-matrix as

\beqa {\cal A}_{1}& = & \frac{\mu_p \pi^{-1/2}}{4}
 \int
 dx_{1}dx_{2}dx_{3}dx_{4} dx_{5}\,
(P_{-}\fsH_{(n)}M_p)^{\alpha\beta}\xi_{1i} \bu ^{\dga} u^{\dde}  (x_{23}x_{24}x_{25}x_{34}x_{35}x_{45})^{-1/4}
\nonumber\\&&\times \bigg(\frac{x_{45}x_{23}}{x_{42}x_{43}x_{52}x_{53}}\bigg)^{1/4}
  \bigg[\frac{C_{\alpha}^{\dde}C_{\beta}^{\dga}} {x_{43}x_{52}}-\frac{C_{\alpha}^{\dga}C_{\beta}^{\dde}} {x_{42}x_{53}}+\frac{1}{2}\frac{(\gamma^\mu C)_{\alpha\beta}\,(\gab_\mu C)^{\dga\dde}} {x_{45}x_{23}}\bigg] I_1
 \Tr(\lam_1\lam_2\lam_3),\labell{125}\eeqa
  $\frac{\mu_p \pi^{-1/2}}{4}$ is normalization constant,and we have used the following definitions
$x_{ij}=x_i-x_j, x_4=z=x+iy, x_5=\bar z=x-iy$ where
\beqa
I_1= \bigg(\frac{ip^i  x_{54}}{x_{14} x_{15}}\bigg)  |x_{12}|^{\alpha'^2 k_1.k_2}|x_{13}|^{\alpha'^2 k_1.k_3}|x_{14}x_{15}|^{\frac{\alpha'^2}{2} k_1.p}|x_{23}|^{\alpha'^2 k_2.k_3}|
x_{24}x_{25}|^{\frac{\alpha'^2}{2} k_2.p}
|x_{34}x_{35}|^{\frac{\alpha'^2}{2} k_3.p}|x_{45}|^{\frac{\alpha'^2}{4}p.D.p},
\nonumber\eeqa
 Notice that now the amplitude is SL(2,R) invariant and for the simplicity we use the satndard  Mandelstam variables
\beqa
s&=&-\frac{\alpha'}{2}(k_1+k_3)^2, \quad t=-\frac{\alpha'}{2}(k_1+k_2)^2, \quad u=-\frac{\alpha'}{2}(k_3+k_2)^2,
\nonumber
\eeqa

We also need to apply a very  distinguished gauge fixing  $(x_1=0,x_2=1,x_3=\infty)$.

After gauge fixing one can derive the following form for the first part of the amplitude in IIA as below

\beqa {\cal A}_{1}& = & \frac{\mu_p \pi^{-1/2}}{4}(P_{-}\fsH_{(n)}M_p)^{\alpha\beta} \bu^{\dga} u^{\dde} (-ip.\xi) \int\int
 dz d\bar z |z|^{2t+2s-2}|1-z|^{2t+2u-1} (z-\bar z)^{-2(t+s+u)+1},
\nonumber\\&&\times   \bigg[\frac{C_{\alpha}^{\dde}C_{\beta}^{\dga}} {1-\bar z}+\frac{C_{\alpha}^{\dga}C_{\beta}^{\dde}} {z-1}-\frac{1}{2}\frac{(\gamma^\mu C)_{\alpha\beta}\,(\gab_\mu C)^{\dga\dde}} {z-\bar z}\bigg]\Tr(\lam_1\lam_2\lam_3)
\labell{amp3q},\eeqa

To have the complete part of the amplitude to all orders in $\alpha'$ one has to take integrations just on the position of the closed string RR, ( see \cite{Fotopoulos:2001pt,Hatefi:2012wj} ). Without any further works we write the all order solution of the first part of our amplitude as below :

\beqa
{\cal A}_{1}^{C \phi \bar\psi \psi}& = & \frac{\mu_p \pi^{-1/2}}{4} (P_{-}\fsH_{(n)}M_p)^{\alpha\beta}\bu^{\dga} u^{\dde} (-ip.\xi)
 \bigg[L_1(C_{\alpha}^{\dde}C_{\beta}^{\dga}-C_{\alpha}^{\dga}C_{\beta}^{\dde}) -\frac{1}{2} L_2 (C_{\alpha}^{\dde}C_{\beta}^{\dga}+C_{\alpha}^{\dga}C_{\beta}^{\dde})\nonumber\\&&-\frac{1}{2}(\gamma^\mu C)_{\alpha\beta}(\gab_\mu C)^{\dga\dde} L_3\bigg] \Tr(\lam_1\lam_2\lam_3)
\labell{amp3q},\eeqa

with
\beqa
L_1&=&(2)^{-2(t+s+u)+1}\pi{\frac{\Gamma(-u+\frac{1}{2})
\Gamma(-s+1)\Gamma(-t+1)\Gamma(-t-s-u+1)}
{\Gamma(-u-t+\frac{3}{2})\Gamma(-t-s+1)\Gamma(-s-u+\frac{3}{2})}},\nonumber\\
L_2&=&(2)^{-2(t+s+u+1)}\pi{\frac{\Gamma(-u+1)
\Gamma(-s+\frac{1}{2})\Gamma(-t+\frac{1}{2})\Gamma(-t-s-u+\frac{3}{2})}
{\Gamma(-u-t+\frac{3}{2})\Gamma(-t-s+1)\Gamma(-s-u+\frac{3}{2})}},\nonumber\\
L_3&=&(2)^{-2(t+s+u)}\pi{\frac{\Gamma(-u)
\Gamma(-s+\frac{1}{2})\Gamma(-t+\frac{1}{2})\Gamma(-t-s-u+\frac{1}{2})}
{\Gamma(-u-t+\frac{1}{2})\Gamma(-t-s+1)\Gamma(-s-u+\frac{1}{2})}},\nonumber\\
\label{Ls}
\eeqa

\vskip.1in

 As it is clear from above just $L_3$ has infinite singularities in $u$-channels and depending on whether $\mu$ takes either the world volume or transverse direction we could have gauge for $n=p$ or scalar poles for $n=p+2$ case. More importantly our S-matrix involves so many contact interactions. Needless to say that the expansion is just done by sending all Mandelstam variables to zero  ( for comprehensive review of the expansions see \cite{Hatefi:2010ik}).

 \vskip.1in

We are about finding the second part of the amplitude.  There is a  subtle issue for this part  as follows. If we substitute the second part of the vertex operator of scalar inside the amplitude  then we observe that one has to derive the correlator of four spin operators with different chiralities and one current in ten dimensions of space time.

Thus the second part of the S-matrix is given by:

 \beqa {\cal A}_{2}^{C \phi \bar\psi \psi}& = & \frac{\mu_p \pi^{-1/2}}{4}\int
 dx_{1}dx_{2}dx_{3}dx_{4} dx_{5}\,
(P_{-}\fsH_{(n)}M_p)^{\alpha\beta}\xi_{1i} (2ik_{1a})\bu^{\dga} u^{\dde} ( x_{23}x_{24}x_{25}x_{34}x_{35} x_{45})^{-1/4} \nonumber\\&&\times
 <:\psi^a\psi^i(x_1):S_{\dga}(x_2):S_{\dde}(x_3):S_{\alpha}(x_4):S_{\beta}(x_5):>
 I
 \Tr(\lam_1\lam_2\lam_3),\labell{1289}\eeqa

 in which
\beqa
I=|x_{12}|^{\alpha'^2 k_1.k_2}|x_{13}|^{\alpha'^2 k_1.k_3}|x_{14}x_{15}|^{\frac{\alpha'^2}{2} k_1.p}|x_{23}|^{\alpha'^2 k_2.k_3}|
x_{24}x_{25}|^{\frac{\alpha'^2}{2} k_2.p}
|x_{34}x_{35}|^{\frac{\alpha'^2}{2} k_3.p}|x_{45}|^{\frac{\alpha'^2}{4}p.D.p},
\nonumber\eeqa

\vskip 0.1in

 In \cite{Hatefi:2013eia} we have explained the method of deriving the correlation function of four spin operators and one current but let us summarize it very briefly once more.
 To do so the first step is to indeed consider the  OPE
 \beqa
 :\psi^{a}\psi^{i}(x_1):S_{\alpha}(x_4):&\sim& -(\gamma^{a}\gamma^{i})_{\alpha}^{\lambda} S_{\lambda}(x_4)x_{14}^{-1}\label{esi1},
 \eeqa
in addition to that we need to substitute the above  relation to our original  correlaor $<:\psi^a\psi^i(x_1):S_{\dga}(x_2):S_{\dde}(x_3):S_{\alpha}(x_4):S_{\beta}(x_5):>$ and use  correlation of   four spin operators  with different chiralities \reef{ham}. The other steps could be easily followed by taking the other different permutations and  finally adding all the terms.  It is also of high importance to mention that the following correlaor must have been considered as well
\beqa
 <:S_{\dga}(x_2):S_{\dde}(x_3):\psi^i(x_1):>&=&2^{-1/2}x_{23}^{-3/4}(x_{12}x_{13})^{-1/2}
(\gamma^{i})_{\dga\dde}.\nonumber
\eeqa

 \vskip.2in

 Finally we should re-construct different combinations of various gamma matrices  and take advantage of the all appendices A.1, A.3, B.3 and in particular section 6 of \cite{Hartl:2010ks}. Indeed we have checked that our amplitude produces all the desired singularities associated to various  channels and more significantly the S-matrix keeps track of   the SL(2,R) invariance. All these points are  showing us that we have obtained the correct and ultimate form of the correlation function of four spin operators (with different chiralities ) and one current. We also emphasize that the integrals are evaluated on the location of RR and a particular gauge fixing (likewise the gauge fixing of the first part of the amplitude)  has been taken.

 \vskip.2in

Final form of the second part of the S-matrix of $RR\bar \Psi \Psi \Phi$ is given as follows:

 \vskip.2in

\beqa {\cal A}_{2}^{C \phi \bar\psi \psi}  & = & \frac{\mu_p \pi^{-1/2}}{4} (P_{-}\fsH_{(n)}M_p)^{\alpha\beta}\xi_{1i}(2ik_{1a}) \bu^{\dga} u^{\dde}  \nonumber\\&&\times\bigg({\cal A}_{21}+{\cal A}_{22}+{\cal A}_{23}+{\cal A}_{24}+{\cal A}_{25}+{\cal A}_{26}+{\cal A}_{27}+{\cal A}_{28}+{\cal A}_{29}\bigg)
\Tr(\lam_1\lam_2\lam_3),\nonumber\eeqa

such that
\beqa {\cal A}_{21} &=&-\frac{1}{2}(\gamma^a \gab^i C)_\alpha{}^{\dga} C_\beta^{\dde}
\bigg[L_4+\frac{s}{2} L_5\bigg]\frac{1}{(-s-u+\frac{1}{2})}
\nonumber\\
 {\cal A}_{22} &=&+\frac{1}{2}(\gamma^a\gab^i C)_\alpha{}^{\dde}C_\beta^{\dga}
\bigg[-L_4+\frac{t}{2} L_5\bigg]\frac{1}{(-t-u+\frac{1}{2})}
\nonumber\\
{\cal A}_{23} &=&\frac{1}{2}(\gamma^a \gab^i C)_\beta{}^{\dga}C_\alpha^{\dde}
\bigg[L_4-\frac{s}{2} L_5\bigg]\frac{1}{(-s-u+\frac{1}{2})}
\nonumber\\
{\cal A}_{24} &=& -\frac{1}{2}(\gamma^a \gab^i C)_\beta{}^{\dde}C_\alpha^{\dga}
\bigg[-L_4-\frac{t}{2} L_5\bigg]\frac{1}{(-t-u+\frac{1}{2})}
\nonumber\\
{\cal A}_{25} &=&-\frac{1}{2}(\gamma^a C)_{\alpha\beta}(\gab^i C)^{\dga\dde}
\bigg[sL_6+\frac{1}{2} L_3\bigg]
\nonumber\\
{\cal A}_{26} &=& -\frac{1}{2}(\gamma^i C)_{\alpha\beta}(\gab^a\,C)^{\dga\dde}
\bigg[-tL_6+\frac{1}{2} L_3\bigg]
\nonumber\\
{\cal A}_{27} &=& L_3\bigg[-\frac{1}{4}(\gamma^a\gab^\lambda C)_\alpha{}^{\dga} (\gamma^i \gab_\lambda C)_\beta{}^{\dde}\bigg]
\nonumber\\
{\cal A}_{28} &=&  L_3\bigg[
  -\frac{1}{4}(\gamma^a \gab^\lambda C)_\alpha{}^{\dde}(\gamma^i \gab_\lambda C)_\beta{}^{\dga} \bigg]
\nonumber\\
{\cal A}_{29} &=& -(-t-s) L_6\bigg[\frac{1}{4}(\gab^a \gamma^i \gab^\lambda C)^{\dga\dde}(\gamma_\lambda C)_{\alpha\beta}\bigg],
\labell{ampc}
\eeqa

where $L_4,L_5,L_6$ are
\beqa
L_4&=&(2)^{-2(t+s+u)}\pi{\frac{\Gamma(-u+1)
\Gamma(-s+\frac{1}{2})\Gamma(-t+\frac{1}{2})\Gamma(-t-s-u+\frac{1}{2})}
{\Gamma(-u-t+\frac{1}{2})\Gamma(-t-s+1)\Gamma(-s-u+\frac{1}{2})}}\nonumber\\
L_5&=&(2)^{-2(t+s+u)+1}\pi{\frac{\Gamma(-u+\frac{1}{2})
\Gamma(-s)\Gamma(-t)\Gamma(-t-s-u+1)}
{\Gamma(-u-t+\frac{1}{2})\Gamma(-t-s+1)\Gamma(-s-u+\frac{1}{2})}}\nonumber\\
L_6&=&(2)^{-2(t+s+u)-1}\pi{\frac{\Gamma(-u+\frac{1}{2})
\Gamma(-s)\Gamma(-t)\Gamma(-t-s-u)}
{\Gamma(-u-t+\frac{1}{2})\Gamma(-t-s+1)\Gamma(-s-u+\frac{1}{2})}}\nonumber\\
\label{Ls24}
\eeqa

\vskip 0.1in

Let us talk about for each part of the amplitude in IIA. Since we are dealing with  massless strings , the expansion is just low energy expansion and if we would send all Mandelstam variables to zero then we would see that $L_1,L_2,L_4$ have no singularities. Indeed these functions are responsible for infinite contact interaction of two fermions with different chiralities and one RR and one massless scalar field. Therefore to be able to work with singular terms and at the level of poles one can easily ignore them. The first part of the amplitude has just infinite $u-$ channel poles (it is obvious from the expansion of $L_3$).

 Thus all terms in the second part of the amplitude carrying $L_4$ coefficient are just contact terms. Notice also to the fact that the second terms of  $ {\cal A}_{21}$ and $ {\cal A}_{23}$ are related to an infinite number of $t-$ channel poles and one has to add them up to re-produce all $t-$ channel poles in field theory side of IIA. More importantly  the second terms of  $ {\cal A}_{22}$ and $ {\cal A}_{24}$ are related to an infinite number of $s-$ channel poles and we need to add them up to re-produce all $s-$ channel poles in field theory side of IIA, accordingly.

 \vskip.2in

Note that all terms accompanying the coefficient of $L_3$ in  $ {\cal A}_{25}$,$ {\cal A}_{26}$ and $ {\cal A}_{27}$ and and $ {\cal A}_{28}$ themselves do include just infinite $u$- channel poles where in field theory we make it clear what kinds of poles (either gauge  or scalar ) can be propagated. In fact the trace and kinematic relation of closed string RR imposes to the amplitude whether gauge or massless scalar must be propagated.

\vskip.2in

 On the other hand the coefficient of the first part of$ {\cal A}_{25}$ which is  $sL_6$ tells us that this part of the amplitude has a double pole in $t$ and $(t+s+u)$ channels and appropriately the coefficient of the first part of$ {\cal A}_{26}$ which is  $-tL_6$ gives  us the information about having a double pole  $s$ and $(t+s+u)$ channels  and finally  $-tL_6$ ($-sL_6$) inside  $ {\cal A}_{29}$ clarifies that we do have a double pole in  $s$ and $(t+s+u)$ ( $t$ and $(t+s+u)$) channels where one has to add them to the other double poles.

  \vskip.2in

 Note that each part of our amplitude is totally anti symmetric with respect to interchanging the fermions and this is a test in favor of our long computations. The other fact which is highly important is that unlike type IIB superstring theory here we do have double poles.

  \vskip.2in

 Neither do we have single nor infinite massless $(t+s+u)$-channel poles in the first part of the amplitude  therefore those corrections that have been derived in IIB  , including infinite corrections to two fermions -two scalar fields and to two fermions, one scalar and one gauge field  are not applicable to  two fermions , one RR and one scalar of IIA at all.

\vskip .1in

 Here we are going to highlight an important comment as follows.

\vskip .1in

 Given above reasons, one can reveal that the closed form of the correlators of this amplitude and the final result of our amplitude is completely different from the same  S-matrix in IIB and it is quite obvious that  one can not obtain the complete form of IIA S-matrix by applying  T-duality to the  results of IIB (\cite{Hatefi:2012ve,Hatefi:2012rx}).

 In fact due to  the presence of the momentum of RR in transverse direction $p^i$ and the fact that we are working in non-compact direction and winding modes are not appeared in RR's vertex operator , we understand that the terms coming with $p^i$ should be derived by just explicit computations and can not be found by duality transformation.
 ( compare  $CA AA$ with $C\phi AA$) . Hence we must apply direct CFT methods even for fermionic amplitudes.

\vskip .1in

 To deal with all contact interactions one needs to make use of low energy expansion  by sending
  ($t,s,u \rightarrow 0$) so that the momentum conservation in world volume holds  $t+s+u=-p^ap_a$ . One can find out $ L_3$ expansion comprehensively as below
\beqa
 L_3&=&-\pi^{3/2}\bigg[\sum_{n=-1}^{\infty}b_n\bigg(\frac{1}{u}(t+s)^{n+1}\bigg)+\sum_{p,n,m=0}^{\infty}e_{p,n,m}u^{p}(st)^{n}(s+t)^m\bigg],
\nonumber\\
-tL_5&=&-\pi^{3/2}\bigg[\sum_{n=-1}^{\infty}b_n\bigg(\frac{1}{s}(t+u)^{n+1}\bigg)+\sum_{p,n,m=0}^{\infty}e_{p,n,m}s^{p}(ut)^{n}(u+t)^m\bigg],
\labell{highcaap}.
\eeqa
where $-sL_5$ could be easily derived from $-tL_5$ by interchanging t to s, with the following coefficients

\beqa
b_{-1}=1,\,b_0=0,\,b_1=\frac{1}{6}\pi^2,\,b_2=2\z(3),
e_{0,0,1}=\frac{1}{3}\pi^2,
\nonumber\\
e_{0,1,0}=2\z(3),e_{1,0,0}=\frac{1}{6}\pi^2,e_{1,0,2}=\frac{19}{60}\pi^4,e_{1,0,1}=6\z(3),
\labell{577}\eeqa
in the next section we first produce  all infinite $u$-channel scalar poles, more importantly we try to come up with some remarks about double poles that are allowed in type IIA , given the complete form of our S-matrix.

\subsection{ New couplings in type IIA and their third order $\alpha'$ corrections  }

 In this section we are going to obtain several important remarks on type IIA superstring theory. First of all by applying direct computations  in string theory (having found our complete S-matrix ), we obtain several new couplings in IIA with their $\alpha'$ corrections, second of all we  want to explore new corrections in IIA.

 Basically we will show that the higher derivative corrections of two fermions-two scalars of  IIB  are not consistent with IIA and in fact not only the coefficients of those corrections of IIA  are different from IIB but also S-matrix imposes us that the general structure of those IIA corrections are entirely different from IIB.

 The third reason is as follows. In IIB we did have the  S-matrix\footnote{ \beqa {\cal A} &=&- \frac{ \alpha'\pi^{-1/2}\mu_p}{(p)!} (\veps^v)^{a_0\cdots a_{p-1}a} H_{a_0\cdots a_{p-1}}\xi_{1i}(2ik_{1a}) \bu_1^{A} (\ga^i )_{AB}u_2^{B}
\Tr\left(\lambda_1\lambda_2\lambda_3\right)\,
\bigg[-2t L_3\bigg]\sum_{m=o}^{\infty} a_{n,m}[t^ns^m+t^ms^n]
 \non
\eeqa.}, since we donot have any of the those terms in IIA , we expect not to have any single correction of two fermions, one scalar and one gauge field of IIA.  Namely the following corrections of IIB
\footnote{
\beqa
&&{\cal L}^{n,m}= \pi^3\alpha'^{n+m+3}T_p\bigg(a_{n,m}\Tr\bigg[\cD_{nm}\left(\bPsi \ga^i D_b\Psi D^a\phi^i F_{ab} \right)+\cD_{nm}\left( \bPsi \ga^i D_b\Psi F_{ab} D^a\phi^i \right)
\nonumber\\&&+h.c \bigg]
+i b_{n,m}\Tr\bigg[\cD'_{nm}\left(\bPsi \ga^i D_b\Psi D^a\phi^i F_{ab} \right)+\cD'_{nm}\left(
 \bPsi \ga^i D_b\Psi F_{ab} D^a\phi^i  \right)+h.c.\bigg]
\bigg).
\non
\eeqa}, are not applicable to IIA.

 Let us start exploring new correcions of type IIA.

 \vskip.1in

 For this section we must consider ${\cal A}_{25} $ and  ${\cal A}_{26} $ ( which are anti symmetric with respect to interchanging t to s), and add them up to ${\cal A}_{29} $. In fact   $(\gamma_\lambda)$ in  $(\gamma_\lambda C)_{\alpha\beta}$ can have just component in transverse direction $\lambda=i$, in such a way that after adding all the terms and extracting the traces , the final form of our S-matrix will be given by the following :

 \beqa {\cal A} &=&- \frac{ ik_{1a}\alpha'\pi^{-1/2}\mu_p}{(p+1)!} (\veps)^{a_0\cdots a_{p}} H^i_{a_0\cdots a_{p}}\xi_{1i}\bu^{\dga} (\ga^a )_{\dga\dde} u^{\dde} (-t+s) L_6
\Tr\left(\lambda_1\lambda_2\lambda_3\right)\,
 \labell{ampcmm},
\eeqa

So it is antisymmetric with respect to interchanging two fermions and it means that the amplitude of one RR, two fermion with the same chirality is zero as we expected, so we just work with  the expansion of $tL_6$  as follows:

\beqa
tL_6&=&\frac{\sqrt{\pi}}{2}\left(\frac{-1}{s(t+s+u)}+ \frac{4\ln(2)}{s}\right.\labell{high2}\\
&&\left.+\left(\frac{\pi^2}{6}-8\ln(2)^2\right)\frac{(s+t+u)}{s}-\frac{\pi^2}{3}\frac{t}{(t+s+u)}+\cdots\right)\nonumber\eeqa

It is clear from the above expansion that unlike IIB in type IIA for one RR and two fermions and one scalar we have several poles, involving double poles and new couplings . In particular we do have  simple poles in
 $(s+t+u)=-p^a p_a$  channels for type IIA.

 The effective field theory of IIA proposed us that  we are having  a double pole in $t+s+u$ and $s$
 channels and they need to be re-constructed for $p+2=n$ case by the following rule.

 \vskip.1in

  The effective field theory has also the following  poles :
\beqa
{\cal A}&=&V_i(C_{p+1},\phi)G_{ij}(\phi)V_j(\phi,\bar\psi_1,\psi_2)G(\psi_2)V(\bar\psi_2,\psi_1,\phi^{(1)})\labell{ampkl5}\eeqa
note that  the off-shell scalar  $\phi $ can be both   $\phi^{(1)}$ and $\phi^{(2)}$. The vertex of $V_i(C_{p+1},\phi)$ could be extracted from
 \footnote{\beqa
(2\pi\alpha')\mu_p\int d^{p+1}\sigma {1\over (p+1)!}
(\veps)^{a_0\cdots a_{p}}\,\Tr\left(\phi^i\right)\,
H^{(p+2)}_{ia_0\cdots a_{p}}, \label{cvb}\eeqa}  to be $V_i(C_{p+1},\phi)=
(2\pi\alpha')\mu_p {1\over (p+1)!}
(\veps)^{a_0\cdots a_{p}}
H^{i}_{a_0\cdots a_{p}}$
 , the scalar propagator is also derived from the kinetic term of the scalar fields to be

\beqa
G_{\alpha\beta}^{ij}(\phi) &=&\frac{-i\delta_{\alpha\beta}\delta^{ij}}{T_p(2\pi\alpha')^2
k^2}=\frac{-i\delta_{\alpha\beta}\delta^{ij}}{T_p(2\pi\alpha')^2
(t+s+u)} \eeqa
where k is the momentum of off-shell scalar field.

Note that  the vertex of $V_j(\phi,\bar\psi_1,\psi_2)$ includes an off-shell scalar field , one on-shell and one off-shell fermion field where each one lives in different brane can be decomposed as follows:

\beqa
V_j(\phi,\bar\psi_1,\psi_2)&=& T_p (2\pi\alpha')\bar u^{\dga} \gamma_{j\dga}\nonumber\eeqa

Fermion propagator could be derived from fermions ' kinetic term and we have to note that this off-shell fermion is the fermion that glued to one external scalar and one on-shell external fermion  so that if we apply the momentum conservation then we can write down this propagator as below:

\beqa
G(\psi) &=&\frac{-i\gamma^a (k_1+k_3)_a}{T_p(2\pi\alpha')
(k_1+k_3)^2}=\frac{-i\gamma^a (k_1+k_3)_a}{T_p(2\pi\alpha')s}\eeqa

Finally we need to find out the vertex of an on-shell scalar and one on -shell/an off-shell fermion  $V(\bar\psi_2,\psi_1,\phi^{(1)})$ , which could be explored from their kinetic terms as below:

\beqa
 V(\bar\psi_2,\psi_1,\phi^{(1)})&=&T_p (2\pi\alpha')\bar u^{\dde} \gamma_{k\dde}\xi_{1k}\nonumber\eeqa

Having replaced all above vertices and make use of this rule ,

\vskip.2in

${\cal A} =V_i(C_{p+1},\phi)G_{ij}(\phi)V_j(\phi,\bar\psi_1,\psi_2)G(\psi_2)V(\bar\psi_2,\psi_1,\phi^{(1)})$
we are able to  obtain the following result:

 \beqa {\cal A} &=&- \frac{ ik_{1a}\mu_p}{s(t+s+u)(p+1)!} (\veps)^{a_0\cdots a_{p}} H^i_{a_0\cdots a_{p}}\xi_{1i}\bu^{\dga} (\ga^a )_{\dga\dde} u^{\dde}
\Tr\left(\lambda_1\lambda_2\lambda_3\right)\,
 \labell{ampcmm66},
\eeqa

 Which is exactly the first term of the expansion so we could precisely produce the double pole. Note that we have also used the $k_{3a}\gamma^a u=0$ which is the equation of motion for fermion field. In order to produce the second term in the expansion of $t L_6$ , one has to consider the following rule with some new couplings.

\beqa
{\cal A}&=&V(C_{p+1},\bar\Psi_1,\Psi_2)G(\Psi_2)V(\bar\Psi_2,\Psi_1,\phi^1),\label{amp54lm9}
\eeqa

where  we have derived fermion propagator  and the $V(\bar\Psi_2,\Psi_1,\phi^1)$ could be once more derived by taking into account fermions' kinetic term as follows
$V(\bar\Psi_2,\Psi_1,\phi^1)=T_p (2\pi\alpha')  u^{\dde} \gamma^j \xi_{1j}$. On the other hand in order to look for the second pole in $tL_6$ expansion, the S-matrix imposes us that  a new WZ coupling in type IIA should be appeared as

 \beqa
\frac{(2\pi\alpha')\mu_p}{(p+1)!}\beta_1^2\int d^{p+1}\sigma
\Tr\left( C_{a_0\cdots a_{p}} \bPsi_1 \ga^l \partial_l \Psi_2 \right) (\veps)^{a_0\cdots a_{p}}
\label{nn2v2}\eeqa

 Now if we choose $\beta_1$  to be

 \beqa
 \beta_1&=& (2ln2/(\pi\alpha'))^{1/2}\nonumber\eeqa
 and make use of the given vertices and the rule ${\cal A}=V(C_{p+1},\bar\Psi_1,\Psi_2)G(\Psi_2)V(\bar\Psi_2,\Psi_1,\phi^1),$ we can obtain the following amplitude in field theory side as below:

 \beqa {\cal A} &=&- \frac{ \alpha' (ln2) ik_{1a}\alpha'\mu_p}{s(p+1)!} (\veps)^{a_0\cdots a_{p}} H^i_{a_0\cdots a_{p}}\xi_{1i}\bu^{\dga} (\ga^a )_{\dga\dde} u^{\dde}
\Tr\left(\lambda_1\lambda_2\lambda_3\right)\,
 \non\eeqa
 which is precisely the second pole of the expansion of  $tL_6$.

 \vskip.2in

 The next question to address is that how we can produce the third term of the expansion of $t L_6$ which is a simple massless fermion pole. The answer is as follows.
\vskip.1in

 The third pole could be looked for by proposing the same rule of \reef{amp54lm9},however, the fermion propagator and the vertex of two fermions and one scalar field do  not receive any correction, therefore to produce that pole , one has to explore the higher derivative corrections to one RR -(p+1) form field and to $\bar\Psi_2$ and
 $\Psi_1$ as below:

 \beqa
\bigg(\frac{\pi^2}{6}-8 ln2^2\bigg)\frac{i (\alpha')^2\mu_p}{(p+1)!}\int d^{p+1}\sigma
\Tr\left( C_{a_0\cdots a_{p}} D^aD_a (\bPsi_1 \ga^l \partial_l \Psi_2) \right) (\veps)^{a_0\cdots a_{p}}
\label{n1n2v2}\eeqa

 Now by taking integration by parts and using the following constraint $s+t+u=-p^ap_a$, we can precisely obtain the correct form of a new  WZ coupling  in the presence of its correction at second order of $\alpha'$  as

 \beqa
 V(C_{p+1},\bar\Psi_1,\Psi_2)&=& \alpha'^2 \mu_p \bigg(\frac{\pi^2}{6}-8 ln2^2\bigg) \frac{1}{(p+1)!}H^i_{a_0\cdots a_{p}} \bar u \gamma_i (t+s+u)(\veps)^{a_0\cdots a_{p}}\label{mmn}\eeqa

 Keeping fixed the fermion propagator and the vertex of two fermion fields and one scalar field and in particular  substituting \reef{mmn} in ${\cal A}=V(C_{p+1},\bar\Psi_1,\Psi_2)G(\Psi_2)V(\bar\Psi_2,\Psi_1,\phi^1),$
 we are actually able to discover the amplitude in field theory as

 \beqa {\cal A} &=&- \bigg(\frac{\pi^2}{6}-8 ln2^2\bigg)\frac{ (t+s+u) ik_{1a}\mu_p}{s(p+1)!} (\veps)^{a_0\cdots a_{p}} H^i_{a_0\cdots a_{p}}\xi_{1i}\bu^{\dga} (\ga^a )_{\dga\dde} u^{\dde}
\Tr\left(\lambda_1\lambda_2\lambda_3\right)\,
 \labell{ampbcg},
\eeqa
 which is exactly the third pole of  $t L_6$ expansion and  we have considered the equations of motion for fermion fields as well.

Consider $-sL_6$ expansion

\beqa
-sL_6&=&\frac{\sqrt{\pi}}{2}\left(\frac{1}{t(t+s+u)}- \frac{4\ln(2)}{t}\right.\labell{high2}\\
&&\left.-\left(\frac{\pi^2}{6}-8\ln(2)^2\right)\frac{(s+t+u)}{t}+\frac{\pi^2}{3}\frac{s}{(t+s+u)}+\cdots\right)\nonumber\eeqa
 by interchanging the fermions $\Psi_1 $ to $\Psi_2$ and $t \leftrightarrow s$, one can precisely show that the first, second and third poles could be produced by the described field theory. On the other hand we now add the last terms of the $tL_6$ and $-s L_6$ to get to final result of the string amplitude for $p+2=n$ case as

 \beqa {\cal A} &=&- \frac{ ik_{1a}\pi^2\mu_p}{3(t+s+u)(p+1)!} (\veps)^{a_0\cdots a_{p}} H^i_{a_0\cdots a_{p}}\xi_{1i}\bu^{\dga} (\ga^a )_{\dga\dde} u^{\dde} (t-s)
\Tr\left(\lambda_1\lambda_2\lambda_3\right)\,
 \labell{ampcmmgy},
\eeqa

As we can see apart from RR 's field strength , the amplitude carries three momenta . One may argue that this simple $(t+s+u)$- pole (which has to be just scalar pole), could be produced by the obtained couplings of two fermions- two scalars of type IIB \cite{Hatefi:2013eia}, however we show that those corrections can not be held in type IIA.

In order to be able to generate the poles in \reef{ampcmmgy}, one must consider the rule as

\beqa
A&=&   V^{\alpha}_i(C_{p+1},\phi) G^{\alpha\beta} _{ij} (\phi) V^{\beta}_{j}(\phi,\bar\Psi,\Psi,\phi_1)\label{esi34}\eeqa

where we have shown that $V^{\alpha}_i(C_{p+1},\phi)$ and scalar propagator will not receive any corrections, thus we need to look for the corrections to $V^{\beta}_{j}(\phi,\bar\Psi,\Psi,\phi_1)$ of type IIA.

It is shown in  \cite{Hatefi:2013eia} that the corrections of two on-shell fermions and an on-shell/ an off-shell scalar of type IIB are

 \beqa
 \frac{T_p (2\pi\alpha')^3}{4}(\bPsi \ga^a D_b\Psi D^a\phi^i D^b\phi_i+D^a\phi^i D^b\phi_i\bPsi \ga^a D_b\Psi  )\label{poli76}  \eeqa
While if we consider \reef{poli76}, extract all the desired orderings  for $V^{\beta}_{j}(\phi,\bar\Psi,\Psi,\phi_1)$
and apply fermions ' equation of motion we find out

\beqa
V_{\beta}^{j}(\phi,\bar\Psi,\Psi,\phi_1)&=&i\frac{T_p (2\pi\alpha')^3}{4} k_{1a}\bar u^{\dga} (\ga^a)_{\dga\dde} u^{\dde} \xi_{1j}\bigg( -\frac{t}{2}+\frac{s}{2}\bigg)\Tr(\lam_1\lam_2\lam_3\lam_\beta)
\label{mio2}\eeqa

Now if we replace \reef{mio2} inside \reef{esi34}, then we obviously reveal that the final result is completely different from the given S-matrix  in \reef{ampcmmgy}. This confirms that the corrections of type IIB do not work for type IIA.

\vskip.1in

The method of finding out the corrections of BPS and non-BPS branes has been comprehensively explined in \cite{Hatefi:2012rx}, let us propose the following corrections of type IIA at third order of $\alpha'$ as follows:

\beqa
&&{\cal L}= \frac{\pi^3}{3}\alpha'^{3}T_p\bigg(\Tr\bigg[\left(\bPsi \ga^a D_b\Psi D^a\phi^{(1i)} D^b\phi_{(1i)} \right)+\left( D^a\phi^{(1i)} D^b\phi_{(1i)}  \bPsi \ga^a D_b\Psi \right)
\nonumber\\&&+h.c \bigg]
-i \Tr\bigg[\left(\bPsi \ga^a D_b\Psi D^a\phi^{(1i)} D^b\phi_{(2i)} \right)+\left(
 D^a\phi^{(1i)} D^b\phi_{(2i)}  \bPsi \ga^a D_b\Psi \right)+h.c.\bigg]
\bigg),\labell{Lnm}
\eeqa

 work out  \reef{Lnm}  and specially  apply standard field theory techniques to the above couplings, then we are able to gain the following vertex
\beqa
V_{\beta}^{j}(\phi,\bar\Psi,\Psi,\phi_1)&=&i\frac{T_p (\pi\alpha')^3}{3} k_{1a}\bar u^{\dga} (\ga^a)_{\dga\dde} u^{\dde} \xi_{1j}\bigg( -t+s\bigg)\Tr(\lam_1\lam_2\lam_3\lam_\beta)
\label{mio2o}\eeqa

 Having replaced scalar propagator $G_{\alpha\beta}^{ij}(\phi) =\frac{-i\delta_{\alpha\beta}\delta^{ij}}{T_p(2\pi\alpha')^2(t+s+u)}$( $k$ is off-shell scalar 's momentum), the vertex of an off-shell scalar and one RR (p+1) form field\footnote{ $V_i(C_{p+1},\phi)=
(2\pi\alpha')\mu_p {1\over (p+1)!}
(\veps)^{a_0\cdots a_{p}}
H^{i}_{a_0\cdots a_{p}}$} and \reef{mio2o}  to   \reef{esi34} we reach to the final result for field theory amplitude as  below:

 \beqa {\cal A} &=&- \frac{ ik_{1a}\pi^2\mu_p}{(t+s+u)(p+1)!} (\veps)^{a_0\cdots a_{p}} H^i_{a_0\cdots a_{p}}\xi_{1i}\bu^{\dga} (\ga^a )_{\dga\dde} u^{\dde} (-t+s) L_6
\Tr\left(\lambda_1\lambda_2\lambda_3\right)\,
 \non\eeqa

 This is precisely the first simple $(t+s+u)$ channel we were looking for.

 Note that to get to the above result we have  taken  momentum conservation  and made use of  fermions' equations of motion. Let us end this section by making an extremely important message about the couplings that appeared in the second line of \reef{Lnm}. Basically we need to apply the on-shell conditions and the fact that we have sent  $p^ap_a$ to zero value , that is, $(t+s+u)=0$ is also largely used. It would be interesting to find out higher derivative corrections of two on-shell fermions and an off-shell/ an on-shell scalar of type IIA to all orders in $\alpha'$ and also to observe whether or not the universal conjecture made and checked for type IIB , works for type IIA superstring theory.

\vskip.01in

Although it is seen that bosonic amplitudes of type IIB and even fermionic amplitudes of type IIB follow a universal conjecture on higher derivative  corrections to all orders of $\alpha'$ \cite{Hatefi:2012rx} (as several checks are made  in \cite{Hatefi:2010ik,Hatefi:2012ve}), it is not clear to us that it so happens for fermionic amplitudes of type IIA, therefore we hope to answer some of these deep questions in favor of exploring all corrections of superstring theory in near future.

\section{ All order  $u-$channel massless scalar poles for $p+2=n$ case }


One may argue that depending on whether $(\lambda^\mu C) _{\alpha\beta}$ in the first part of the amplitude is carried on world volume or transverse we could have gauge or scalar field $u-$ channel pole, however the coefficient of $(p.\xi)$ comes from interaction of closed string RR with a scalar where the scalar comes from Taylor expansion. This clarifies that the first part of the amplitude can have just scalar pole and no massless gauge pole is allowed.
We should highlight the fact that all contact interactions of the second term of  ${\cal A}_{25}$ are overlooked as in this section we would like to obtain just all massless scalar singularities. However, there are already several literatures to deal with contact terms in string theory \cite{Hatefi:2012wj,Hatefi:2012ve}.

Thus $\mu$ should have been in transverse direction for the first part of our amplitude.   Therefore one might extract the related traces to actually write down
all infinite massless scalar u-channel poles as below:

\beqa
{\cal A}&=& \frac{-\alpha'\mu_p i(p.\xi)\pi}{(p+1)!} \sum_{n=-1}^{\infty}b_n\bigg(\frac{1}{u}(t+s)^{n+1}\bigg) \bu^{\dga} (\ga_j )_{\dga\dde}  u^{\dde}
(\veps)^{a_0\cdots a_{p}}H^{j}_{a_0\cdots a_{p}} \Tr(\lam_1\lam_2\lam_3).
\labell{ampbb}\eeqa

A normalization constant $\frac{\mu_p \pi^{-1/2}}{4}$ is also used.

\vskip.1in

We have already shown that  $T_p (2\pi\alpha')\Tr\bigg(\bPsi\ga^aD_a\Psi\bigg)$ (fermion fields' kinetic term ) does not obtain any correction and indeed all the kinetic terms inside the DBI action have no correction\footnote{ see their fixed form as

\beqa
-T_p (2\pi\alpha')\Tr\left(\frac{(2\pi\alpha')}{2} D_a \phi^i D^a\phi_i-\frac{(2\pi\alpha')}{4}F_{ab}F^{ba}-\bPsi\ga^aD_a\Psi\right)\labell{kin}
\eeqa} \cite{Hatefi:2010ik,Hatefi:2012wj,Hatefi:2012ve}.  One off-shell scalar and two on-shell fermions $V_{\beta}^{j}(\phi,\bPsi,\Psi)$  could be found by extracting the connection or commutator inside the kinetic term of fermions
 \beqa
 V^{\beta}_{j}(\bPsi,\Psi,\phi)&=&T_p(2\pi\alpha')\bu^{\dga}\ga_{j\dga\dde}u^{\dde} \Tr(\lam_2\lam_3\lam^\beta),
 \label{rr3}\eeqa

In field theory we need to work out either pull-back ways, Taylor-expansion (see \cite{Hatefi:2012wj} ) or deal with new Wess-Zumino couplings which are all order corrections to Myers action \cite{Myers:1999ps}.

To be able to produce all singularities,  we also need to find out the scalar propagator for which it has been fixed and received no correction.  It should be extracted from scalar field's kinetic term $-T_p  \Tr\left(\frac{(2\pi\alpha')^2}{2} D_a \phi^i D^a\phi_i\right)$ as follows:

 \beqa
 G_{\alpha\beta}^{ij}(\phi) &=&\frac{-i\delta_{\alpha\beta}\delta^{ij}}{T_p(2\pi\alpha')^2 u},
 \label{rr3}\eeqa

The important point here is that the connections must be dropped out.

We also need to employ all order corrections to Taylor expansion of one RR, an off-shell and an on-shell scalar field \footnote{ \beqa
i\frac{(2\pi\alpha')^2\mu_p}{2!(p+1)!}\int d^{p+1}\sigma
(\veps^v)^{a_0\cdots a_{p}}\,\sum_{n=-1}^{\infty}b_n(\alpha')^{n+1}\Tr\left(\partial_i\partial_j C^{(p+1)}_{a_0\cdots a_{p}}
\partial^{a_0}\cdots \partial^{a_n}\phi^i \partial_{a_0}\cdots \partial_{a_n}\phi^j\right).\,\label{rr55}\eeqa
}where these corrections have been derived in \cite{Hatefi:2013eia}. If we apply field theory techniques to those corrections, then the vertex of a RR $p+1$ form field, one on-shell/one off-shell scalar field $V_{\alpha}^{i}(C_{p+1},\phi_1,\phi)$ to all orders in $\alpha'$ can be extracted  as follows:


\beqa
 V_{\alpha}^{i}(C_{p+1},\phi_1,\phi)=
 i\frac{-ip^i H^{j}_{a_0\cdots a_{p}} \xi_{1j}\Tr(\lam^\alpha\lam_1)(2\pi\alpha')^2\mu_p}{2!(p+1)!}
(\veps)^{a_0\cdots a_{p}}\sum_{n=-1}^{\infty}b_n (-\alpha'(k_1.k_2+k_1.k_3))^{n+1}\,
 \label{rr211}\eeqa

Now if we replace the above equations inside the following rule
\beqa
{\cal A} &=& V_{\alpha}^{i}(C_{p+1},\phi_1,\phi)G_{\alpha\beta}^{ij}(\phi) V_{\beta}^{j}(\phi,\bPsi,\Psi)\nonumber\eeqa

 then we are able to precisely generate all massless $u-$ channel scalar poles in type IIA. Therefore not only RR field  induced all order corrections to an on-shell/ an off-shell scalar field  in type IIB string theory \cite{Hatefi:2013eia} but also it imposed the same corrections to type IIA string theory and from this point of view it is a universal phenomenon which does work for both BPS and non-BPS actions \cite{Hatefi:2010ik,Hatefi:2012wj,Hatefi:2012ve,Hatefi:2012rx,Hatefi:2013mwa}.
\vskip.2in

We now turn to see whether or not we do have an infinite number of $u$- channel gauge poles in type IIA .

\vskip.1in

One may think due to the following trace in the  the second term of  ${\cal A}_{25}$
\beqa
(P_{-}\fsH_{(n)}M_p)^{\alpha\beta}(\ga^a C)_{\alpha\beta}&=&\frac{32 }{2(p)!} (\veps)^{a_0\cdots a_{p-1}a}H_{a_0\cdots a_{p-1}} \eeqa

and because of $(\lambda^a C) _{\alpha\beta}$ we do have infinite u-channel gauge poles for IIA. However, if we add the second terms of
$ {\cal A}_{25}$, $ {\cal A}_{26}$ and both $ {\cal A}_{27}$ and $ {\cal A}_{26}$ and also use various identities then one can clearly observe that
unlike the closed form of the correlators  $<V_{C}V_{\bar\psi}V_{\psi} V_{\phi}>$ of type IIB, here for type IIA we have no longer any massless gauge poles left over.

\vskip.1in

Let us provide several physical reasons in favor of the above discussion.
The first reason is as follows. If we consider the second term of $ {\cal A}_{26}$, then the trace imposes us that this part is not vanished for $p+2=n$ case and  then just RR $(C_{p+1})$ can have non zero value, while if we want to have gauge pole we need to have $C_{p-1}$ form field to have non zero Chern-Simons coupling $
i\frac{(2\pi\alpha')^2\mu_p}{p!}\int d^{p+1}\sigma
\Tr\left(\partial_i C_{(p-1)}\wedge F \phi^i\right)$. The second reason is that we have shown in type IIB
\cite{Hatefi:2013eia} that all gauge poles should be produced by the following terms

\beqa
{\cal A}_{24} &=& \frac{\alpha'^2\mu_p \pi\xi_{1i}(ik_{1a}) }{p!} \bu^{A} (\ga_b)_{AB}  u^{B}  \sum_{n=-1}^{\infty}b_n \frac{1}{u}(t+s)^{n+1}
(\veps)^{a_0\cdots a_{p-2}ab}H^{i}_{a_0\cdots a_{p-2}}\Tr(\lam_1\lam_2\lam_3)
 \non\eeqa
meanwhile if we take the second part of $ {\cal A}_{25}$ then we are left with $\bu^{\dga} (\ga_i)_{\dga\dde}  u^{\dde}$ which is inconsistent with above poles. The third reason is that for $\lambda=i$ ($\lambda=b$) in $ {\cal A}_{28}$  we would have $C_p$  ($C_{p-2}$)  form field whereas both
 $C_p$  ($C_{p-2}$) can be interacted neither by field strength of gauge field nor by covariant derivative of scalar field to give rise all $p+1$ indices of the world volume so Chern-Simons terms can not be held for these two relevant cases.

 \vskip.2in

 Finally the last reason is that we can not expect to be able to produce  any contact terms by $(\veps)^{a_0\cdots a_{p-2}} C_{a_0\cdots a_{p-2}} \bPsi \gamma^a D_a \Psi  $
 coupling as sum of the indices can not cover all $p+1$ world volume directions. Hence there are  no any gauge poles for this amplitude of type IIA.
 Therefore neither the rule for type IIB \footnote{$ V_{\alpha}^{a}(C_{p-1},\phi_1,A)G_{\alpha\beta}^{ab}(A) V_{\beta}^{b}(A,\bPsi_1,\Psi_2) $} nor the following corrections of type IIB \footnote{
 \beqa
i\frac{(2\pi\alpha')^2\mu_p}{p!}\int d^{p+1}\sigma
 \,\sum_{n=-1}^{\infty}b_n(\alpha')^{n+1}\Tr\left(\partial_i C_{(p-1)} \wedge
\partial^{a_0}\cdots \partial^{a_n}F \partial_{a_0}\cdots \partial_{a_n}\phi^i\right).\,
 \label{mm55}\eeqa}
  will be held for type IIA.

\section{An Infinite number of  fermion poles for $p+2=n$ case}
It is discussed in  \cite{Hatefi:2013eia} that due to saturation of super ghost charge to the S-matrix , we can not have any gauge/ scalar , tachyon  or graviton/ closed string poles. Therefore to be able to construct all $t,s$ -channel poles , one must consider the fermion poles.

All the second terms of $A_{21},A_{23}$ ($A_{22},A_{24}$) are exactly related to an infinite number of t(s)- channel poles of our S-matrix, meanwhile all the first terms of  $A_{21},A_{22},A_{23},A_{24}$, are corresponded to various contact interactions for different $p,n$ cases .

We first re-consider all s(t)- channel fermion poles of our S-matrix as follows:

\beqa
{\cal A}&=&  \frac{\mu_p \pi^{-1/2}}{4} (P_{-}\fsH_{(n)}M_p)^{\alpha\beta}\xi_{1i}(2ik_{1a}) \bu^{\dga} u^{\dde}  \bigg\{\frac{-sL_5}{4(-s-u+\frac{1}{2})}\bigg[(\gamma^a \gab^i C)_\alpha{}^{\dga} C_\beta^{\dde}
+(\gamma^a \gab^i C)_\beta{}^{\dga}C_\alpha^{\dde}\bigg]
\nonumber\\&&
+\frac{tL_5}{4(-t-u+\frac{1}{2})}\bigg[(\gamma^a\gab^i C)_\alpha{}^{\dde}C_\beta^{\dga}+(\gamma^a \gab^i C)_\beta{}^{\dde}C_\alpha^{\dga}\bigg]\bigg\}
\label{pol1}\eeqa
As we can see the amplitude is antisymmetric with respect to interchanging the fermions.
Now one could replace the expansions of $(-sL_5) $ and $(tL_5) $, extract the related traces and simplify the amplitude more. By applying these points, we are able to write down
all $t$,($s$)-channel fermion poles of the string amplitude , however,  due to antisymmetric property of fermion poles, here we are just going to write down all infinite s-channel fermion poles and obviously at the end one could produce an infinite number of t-channel fermion poles by just changing the fermions so all infinite s-chnnel fermion poles in string theory should be given by

\vskip.1in

\beqa
{\cal A} &=& \frac{\mu_p \pi\xi_{1i}(\alpha')^2 ik_{1a} }{(p+1)!} \bu^{\dga} (\ga^a)_{\dga\dde}  u^{\dde}  \sum_{n=-1}^{\infty}b_n  \frac{1}{s}(u+t)^{n+1}
(\veps)^{a_0\cdots a_{p}}H^{i}_{a_0\cdots a_{p}}\Tr(\lam_1\lam_2\lam_3).
 \label{nmm1}\eeqa

Notice that we have overlooked all the terms carrying the coefficients of   $L_4$ because they are just contact interaction.

\vskip.1in

It is discussed in detail that to be able to produce all infinite s- channel fermion poles one has to use a particular rule \footnote{
\beqa
{\cal A}&=& V_{\alpha}(C_{p+1}, \bar\Psi_2,\Psi)G_{\alpha\beta}(\Psi) V_{\beta}(\bar\Psi,\Psi_3,\phi_1),\labell{amp560}
\eeqa
}

It is crucial to mention that the kinetic term of fermion fields is indeed fixed , so there is no correction to this term and if we use that , then one can derive the fermion propagator as below:

 \beqa
 G_{\alpha\beta}(\psi) &=&\frac{-i\delta_{\alpha\beta}\ga^b(k_1+k_3)_b}{T_p(2\pi\alpha') s}.
 \label{mm55}\eeqa

where the momentum conservation in world volume has been used, the next step is to extract covariant derivative of
fermion field

 \beqa
 D^i\psi= \partial^i\psi-i[\phi^i,\psi]\nonumber\eeqa

 and apply standard field theory  techniques to write  the needed vertex of an on-shell/ an off-shell fermion and one on-shell scalar field   $V_{\beta}(\bar\Psi,\Psi_3,\phi_1)$

 \beqa
 V_{\beta}(\bar\Psi,\Psi_3,\phi_1)&=&T_p (2\pi\alpha')u^{\dde}\gamma^{j}_{\dde}\xi_{1j}\Tr(\lam_3\lam_1\lam^\beta)
 \label{mm5k5}\eeqa

We have discussed about Chern-Simons terms and Wess-Zumino terms including  RR and an arbitrary scalar or gauge fields and one can generalize those couplings/actions to their supersymmetrized version  as follows\footnote{ notice that we write above coupling in such a way that the integrations should be taken over the whole world volume space , that is, total indices should cover all $p+1$ direction of space -time  and one must keep in mind fermions'  equations of motion  $(\fsk_{2a}\bu=\fsk_{3a}u=0 )$ .}

 \beqa
i\frac{(2\pi\alpha')\mu_p}{(p+1)!}\int d^{p+1}\sigma
\Tr\left( C_{a_0\cdots a_{p}} \bPsi \ga^j \partial_j \Psi \right) (\veps^v)^{a_0\cdots a_{p}}.\,
\label{nn2v2}\eeqa

In addition to that , the vertex operator of one on-shell closed string Ramond-Ramond $(p+1)-$ form and an on-shell/an off-shell fermion field is also needed ($V_{\alpha}(C_{p+1},\bar \Psi_2,\Psi)$) which must be found by using  \reef{nn3m3} \footnote{\beqa
 V_{\alpha}(C_{p+1},\bar \Psi_2,\Psi)&=&i\frac{(2\pi\alpha')\mu_p}{(p+1)!}
 H^i_{a_0\cdots a_{p}} \ga^i_{\dga} \bu^{\dga}
(\veps)^{a_0\cdots a_{p}}\,\Tr\left(\lambda_2\lambda^{\alpha}\right).\,
 \label{nn3m3}\eeqa}.

Now given the above rule and vertices, one can feasibly show that the first simple s-channel fermion pole is actually produced in field theory , however, as we might understand , there are an infinite fermion poles. Interesting point is that the above rule holds for an infinite number of fermion poles as well where one just needs to impose  an infinite number of higher derivative corrections to $V_{\alpha}(C_{p+1},\bar \Psi_2,\Psi)$  as below

\beqa
i\frac{(2\pi\alpha')\mu_p}{(p+1)!}\int d^{p+1}\sigma
\sum_{n=-1}^{\infty}b_n(\alpha')^{n+1}\Tr\left( C_{a_0\cdots a_{p}} \partial^{a_0}\cdots \partial^{a_n} \bPsi \ga^i \partial_{a_0}\cdots \partial_{a_n} \partial_i \Psi \right)\,(\veps)^{a_0\cdots a_{p}}.
\label{nn5kl5}\eeqa

It is important to mention that neither simple fermion pole nor two fermion one scalar vertex received any correction as they have been obtained from the fixed kinetic term of open strings.

In  \reef{nn5kl5}, one could replace the partial derivative by covariant derivative and ignore the connection terms , however, in order to see whether those connections can be still kept in the covariant derivative, one must go to six or higher point function which are beyond our access of this paper.

 Let us write down the complete and all order $\alpha'$ corrections of $V_{\alpha}(C_{p+1},\Psi_2,\bar \Psi)$ as follows

 \beqa
 V_{\alpha}(C_{p+1},\bar \Psi_2,\Psi)&=&i\frac{(2\pi\alpha')\mu_p}{(p+1)!} H^i_{a_0\cdots a_{p}} \ga^i_{\dga}
 \bu^{\dga}
(\veps)^{a_0\cdots a_{p}}\,\Tr\left(\lambda_2\lambda^{\alpha}\right)\,\sum_{n=-1}^{\infty}b_n (\alpha'(k_3.k_2+k_1.k_2))^{n+1}.
 \non\eeqa
fermions' equations of motion were also used. Having used the above all order $\alpha'$  vertex and keeping fixed the other vertices, we are precisely able to actually  generalize and construct all order poles in field theory. This obviously clarifies that the presence of just a closed string RR $p+1$ form field has proposed infinite corrections to two fermions in both IIA and IIB and for this specific case corrections remain invariant .

It is very worthwhile to mention that this idea (producing an infinite number of poles with involving RR field by applying all infinite higher derivative corrections to open strings ) is an important result where it is importance will be known in various higher point BPS and non-BPS  branes  so that certainly we do not need to know  any knowledge about world-sheet integrals and  indeed all singularities of higher point functions can be easily derived.

\vskip.2in

Now  having compared field theory coupling with all infinite S-matrix elements , we are going to construct all order $\alpha'$ corrections (without any ambiguities) to two fermions and one scalar and one closed string RR $p+1$ form field in type IIA  and finally fix its coefficient. We first replace the desired expansions inside the S-matrix and consider the following coupling

 \beqa
i\frac{(2\pi\alpha')^2\mu_p}{(p+1)!}\int d^{p+1}\sigma
\Tr\left( \partial_{i} C_{a_0\cdots a_{p}} \bPsi \ga^j \partial_j \Psi \phi^i\right) (\veps)^{a_0\cdots a_{p}}.\,
\label{nn2v2}\eeqa

Comparing with S-matrix elements, one could find out all order $\alpha'$ corrections of the above coupling in type IIA as follows:

\beqa
&&\sum_{p,n,m=0}^{\infty}e_{p,n,m} (\alpha')^{2n+m-2}(\frac{\alpha'}{2})^{p}\frac{(2\pi\alpha')^2\mu_p}
{\pi(p+1)!}\int d^{p+1}\sigma
 \Tr \bigg(\partial_{i} C_{a_0\cdots a_{p}} D_{a_1}\cdots D_{a_n}D_{a_{n+1}}\cdots D_{a_{2n}}\nonumber\\&&\times D^{a_1}\cdots D^{a_m}\bPsi \ga^j  (D^aD_a)^p D_{a_1}\cdots D_{a_m}(\partial_j \Psi D^{a_1}\cdots D^{a_n} D^{a_{n+1}}\cdots D^{a_{2n}} \phi^i )\bigg) \label{newc22}\eeqa

Note that in the above for all order corrections, the connection terms must be dropped out, but we expect that those commutators will be held in higher point functions as well. Note that the partial derivative on fermion in transverse direction must be moved to RR by taking integration by parts.

\section{Conclusions}

We have used the direct CFT methods to actually find out the complete and closed form of the amplitude of two fermions with different chirality (in type IIA superstring theory), a massless scalar field and one closed string Ramond-Ramond field . Indeed due to various reasons it is very important to have the entire correlators that appeared in $<V_{C}V_{\bar\psi}^{\dga}V_{\psi}^{\dde} V_{\phi}>$ of IIA. Probably one can mention that the most important reason to have the complete result of the S-matrix is to find out $\alpha'$ corrections in different theories.

\vskip.2in

We have shown that there are infinite  massless scalar poles  $u$-channels  of type IIA for $p+2=n$ case . Unlike the same amplitude of $<V_{C}V_{\bar\psi} V_{\psi} V_{\phi}>$ of type IIB, here in IIA we have explicitly shown that there is no any gauge pole. All infinite fermionic poles at $t,s$- channels with their all order $\alpha'$ corrections are also explored.

\vskip.2in

  Unlike the IIB amplitude , explicit computations gave rise the fact that in IIA the couplings between two fermions/ one gauge and one scalar and their corrections are not existed any more.

The explicit form of the amplitude clearly imposed us that several new couplings should accompany the effective field theory of type IIA while those couplings did not appear in type IIB.

The other important result is that the first simple $(t+s+u)-$ channel scalar pole has clarified for us that the new forms of the higher derivative corrections of two fermions (with different chirality) and an on-shell/ an off-shell scalar should be written. In fact in order to reconstruct
the $(t+s+u)-$ channel scalar pole, one has to explore the new corrections at order of $\alpha'^3$ ( see \reef{Lnm}) where the general structure of these corrections and most significantly their coefficient is different from IIB corrections and it is a very crucial fact in favor of  having applied direct CFT methods.

\section*{Acknowledgments}

I would like to thank E.Witten, Joe. Polchinski, L.Alvarez-Gaume, E.Martinec, N.Arkani-Hamed ,I.Klebanov,J.Schwarz, C.Vafa, J.Maldacena, L.Rastelli,W.Siegel, L.Bonora, K.S.Narain and N.Lambert  for valuable discussions/comments and for their great remarks. It is true that without their knowledge I was not able to  finish up the last stages of this paper. It is also a fact that , almost all of the computations of this paper were done during my visits to Simons center for geometry and physics at Stony Brook, Institute for advanced study at Princeton , NJ, University of California at Berkeley, KITP, Center for the Fundamental Laws of Nature
at Harvard University and at Caltech. I would like to thank great hospitality of those universities/institutes and in particular acknowledge Nima Arkani-Hamed, P.Horava, C.Vafa, O.Ganor, Mike Douglas and  J.Schwarz.

\end{document}